\documentclass[sigconf,authorversion]{acmart}
\usepackage{todonotes}
\usepackage{float}
\usepackage[breakable]{tcolorbox}
\usepackage{tcolorbox}
\usepackage{booktabs}
\usepackage{enumitem}
\usepackage[relative,overlay]{textpos}
\usepackage[strict]{changepage}
\usepackage{framed}

\definecolor{formalshade}{rgb}{0.51,0.51,0.51}
\definecolor{formalshade-bloom}{rgb}{0.42,0.44,0.48}
\definecolor{formalshade-course}{rgb}{0.22,0.74,0.82}
\definecolor{formalshade-qtype}{rgb}{0.6,0.91,0.52}
\definecolor{formalshade-output}{rgb}{0.6,0.64,0.71}
\definecolor{formalshade-cname}{HTML}{2CA0E1}
\definecolor{formalshade-lo}{HTML}{317ECF}
\definecolor{formalshade-qtype2}{HTML}{FF7A46}
\definecolor{formalshade-unit}{HTML}{43D6AC}
\definecolor{formalshade-unit-light}{HTML}{D5F6ED}
\definecolor{formalshade-btlevel}{HTML}{F9F473}
\definecolor{formalshadelight}{rgb}{0.95,0.95,0.95}
\definecolor{data}{rgb}{0.70,0.24,0.81}

\newenvironment{formal}[1][formalshade]{%
  \def\FrameCommand{%
    \hspace{1pt}%
    {\color{#1}\vrule width 5pt}%
    {\color{formalshadelight}\vrule width 4pt}%
    \colorbox{formalshadelight}%
  }%
  \MakeFramed{\advance\hsize-\width\FrameRestore}%
  \noindent\hspace{-4.55pt}
  \begin{adjustwidth}{}{7pt}%
  \vspace{2pt}\vspace{2pt}%
}
{%
  \vspace{2pt}\end{adjustwidth}\endMakeFramed%
}

\AtBeginDocument{%
  \providecommand\BibTeX{{%
    \normalfont B\kern-0.5em{\scshape i\kern-0.25em b}\kern-0.8em\TeX}}}

\copyrightyear{2024}
\acmYear{2024}
\setcopyright{rightsretained}
\acmConference[ACE 2024]{Australian Computing Education Conference}{January 29-February 2, 2024}{Sydney, NSW, Australia}
\acmBooktitle{Australian Computing Education Conference (ACE 2024), January 29-February 2, 2024, Sydney, NSW, Australia}
\acmDOI{10.1145/3636243.3636256}
\acmISBN{979-8-4007-1619-5/24/01}

\begin{document}

\title{A Comparative Study of AI-Generated (GPT-4) and Human-crafted MCQs in Programming Education}

\author{Jacob Doughty}
\authornote{Both authors contributed equally to this research.}
\email{jadought@andrew.cmu.edu}
\affiliation{%
  \institution{Carnegie Mellon University}
  \city{Pittsburgh}
  \state{PA}
  \country{USA}
}

\author{Zipiao Wan}
\authornotemark[1]
\email{zwan@andrew.cmu.edu}
\affiliation{%
  \institution{Carnegie Mellon University}
  \city{Pittsburgh}
  \state{PA}
  \country{USA}
}

\author{Anishka Bompelli}
\affiliation{%
  \institution{Carnegie Mellon University}
  \city{Pittsburgh}
  \state{PA}
  \country{USA}
}

\author{Jubahed Qayum}
\affiliation{%
  \institution{Carnegie Mellon University}
  \city{Pittsburgh}
  \state{PA}
  \country{USA}
}

\author{Taozhi Wang}
\affiliation{%
  \institution{Carnegie Mellon University}
  \city{Pittsburgh}
  \state{PA}
  \country{USA}
}

\author{Juran Zhang}
\affiliation{%
  \institution{Carnegie Mellon University}
  \city{Pittsburgh}
  \state{PA}
  \country{USA}
}

\author{Yujia Zheng}
\affiliation{%
  \institution{Carnegie Mellon University}
  \city{Pittsburgh}
  \state{PA}
  \country{USA}
}

\author{Aidan Doyle}
\affiliation{%
  \institution{Carnegie Mellon University}
  \city{Pittsburgh}
  \state{PA}
  \country{USA}
}

\author{Pragnya Sridhar}
\affiliation{%
  \institution{Carnegie Mellon University}
  \city{Pittsburgh}
  \state{PA}
  \country{USA}
}

\author{Arav Agarwal}
\affiliation{%
  \institution{Carnegie Mellon University}
  \city{Pittsburgh}
  \state{PA}
  \country{USA}
}

\author{Christopher Bogart}
\affiliation{%
  \institution{Carnegie Mellon University}
  \city{Pittsburgh}
  \state{PA}
  \country{USA}
}

\author{Eric Keylor}
\affiliation{%
  \institution{Carnegie Mellon University}
  \city{Pittsburgh}
  \state{PA}
  \country{USA}
}

\author{Can Kultur}
\affiliation{%
  \institution{Carnegie Mellon University}
  \city{Pittsburgh}
  \state{PA}
  \country{USA}
}

\author{Jaromir Savelka}
\affiliation{%
  \institution{Carnegie Mellon University}
  \city{Pittsburgh}
  \state{PA}
  \country{USA}
}

\author{Majd Sakr}
\affiliation{%
  \institution{Carnegie Mellon University}
  \city{Pittsburgh}
  \state{PA}
  \country{USA}
}

\renewcommand{\shortauthors}{Doughty and Wan, et al.}

\begin{abstract}
There is a constant need for educators to develop and maintain effective up-to-date assessments. While there is a growing body of research in computing education on utilizing large language models~(LLMs) in generation and engagement with coding exercises, the use of LLMs for generating programming MCQs has not been extensively explored. We analyzed the capability of GPT-4 to produce multiple-choice questions (MCQs) aligned with specific learning objectives (LOs) from Python programming classes in higher education. Specifically, we developed an LLM-powered (GPT-4) system for generation of MCQs from high-level course context and module-level LOs. We evaluated 651 LLM-generated and 449 human-crafted MCQs aligned to 246 LOs from 6 Python courses. We found that GPT-4 was capable of producing MCQs with clear language, a single correct choice, and high-quality distractors. We also observed that the generated MCQs appeared to be well-aligned with the LOs. Our findings can be leveraged by educators wishing to take advantage of the state-of-the-art generative models to support MCQ authoring efforts.
\end{abstract}





\maketitle

\section{Introduction}

Multiple-choice question (MCQ) tests are one of the most popular types of assessment in education \cite{butler2018multiple,towns2014guide}. However, crafting high-quality MCQs that accurately target the intended learning objectives (LOs) requires valuable expertise, is time-consuming and, hence, expensive. This is especially true in technical domains such as computing education where developing effective MCQs poses distinct challenges, e.g., those related to inclusion of pieces of computer code. With changes in technology, growing interest in programming education, and low barriers for students to share past assessments, the demand on instructors to author novel high-quality MCQs has never been higher. Recent developments in large language models (LLMs), such as generative pre-trained transformers (GPT), show tremendous potential for addressing this challenge. Leveraging the capabilities of LLMs, educators could potentially (semi-)automate the generation of MCQ assessments.

We developed a novel LLM-based (GPT-4) pipeline to automate generation of MCQs for higher-education Python programming courses. The novelty of our approach lies in making use of a high-level course context and detailed module-level LOs for producing well-formed high-quality MCQs that use clear language, have plausible distractors, and are well-aligned with the LOs. Since understanding the quality and effectiveness of automatically generated MCQs is of utmost importance we performed a rigorous evaluation of 651 automatically generated and 449 human-crafted questions. If high-quality automated MCQ generation proves feasible it could significantly reduce the time and effort educators currently spend on developing assessments. 


To investigate if and how GPT-4 could generate high quality MCQ assessments for higher education programming courses, we analyzed the following research questions:

\begin{itemize}
    \item[\textbf{RQ1:}] To what degree do the generated MCQs meet typical quality requirements? Specifically, do they:
    \begin{itemize}
        \item[(i)] provide sufficient information in clear language;
        \item[(ii)] have a single correct answer with
        \item[(iii)] high-quality distractors; and
        \item[(iv)] contain syntactically and logically correct code?
    \end{itemize}
    \item[\textbf{RQ2:}] How well are the generated MCQs aligned with the specified module-level LOs?
\end{itemize}

By carrying out this work, we provide the following contributions to the computing education research community. To our best knowledge, this is:

\begin{itemize}
    \item[\textbf{C1:}] One of the first studies employing and evaluating LLMs in automatic generation of MCQs for programming classes.
    \item[\textbf{C2:}] One of the first studies generating MCQs not from short pieces of course materials, but from LOs.
    \item[\textbf{C3:}] One of the most extensive (1,100 MCQs) and detailed evaluations of generated MCQs including alignment with LOs.
\end{itemize}

\section{Related Work}
\label{sec:related_work}
As manually constructing MCQs requires significant effort, researchers have focused heavily on the task of automated question generation. Most work in this field focuses on a specific MCQ element, i.e., the \emph{stem} (question), the \emph{key} (correct answer), and the \emph{distractors} (incorrect options). The most widely researched task is question answering (QA) \cite{allam2012question,soares2020literature}; this is equivalent to key generation, although most work in question answering do not place themselves within this context.

The automatic generation of stems is related to generating free-form questions (QG). Initially, these concentrated on reading comprehension tasks, relying on both traditional NLP methods \cite{heilman2011automatic} and neural networks \cite{hochreiter1997long} to yield meaningful questions \cite{duan2017question}. More recent systems rely on sequence models \cite{wang2018qg} or attention-based methods like LLMs \cite{du2017learning,cho2019contrastive,lopez2020transformer,chan2019recurrent} (GPT-2 \cite{radford2019language}, BERT \cite{devlin2018bert}). Additionally, there have been attempts to tackle QA and QG tasks jointly \cite{tang2017question}. Kurdi et al. provide a comprehensive overview of automated free response and MCQ generation for educational purposes \cite{kurdi2020systematic}. 

\begin{table*}[!h!t]
    \centering
    \begin{tabular}{l|r|r|rrrrrrrr}
    \cmidrule{1-11}
                                               &         &          & \multicolumn{6}{c}{LO Bloom's Taxonomy Levels} \\
    Course Name                                & Modules & MCQs     & RMB & UND & APP & ANL & EVL & CRT & N/A & Overall \\
    \midrule
    Practical Programming with Python          &       8 & 194      & 14  & 31  & 20  & 17  & 2   & 29  & 10 & 123 \\
    Get Started with Python                    &       5 & -        & 8   & 24  & 9   & 4   & 1   & 5   & 4 & 55 \\
    Python Essentials -- Part 1 (Basics)\footnote{OpenEDG: Python Essentials - Part 1 (Basics). Available at: \url{https://edube.org/study/pe1} [Accessed 2023-07-30]}       &       4 & 129      & -   &  7  &  -  &  -  &  -  & -   & 14 & 21  \\
    Python Essentials -- Part 2 (Intermediate) &       4 & 126      &  1  & 1  &  4  & 2   & -   &  7  & 5 & 20   \\
    Introduction to Data Science in Python     &       4 & -        &  3  &  2 &  8  &  -  & -   &  1  &  -  & 14  \\
    Python and Pandas for Data Engineering     &       4 & -        &  1  &  1  & 2   &  -  &   - &  9  & -  & 13  \\
    \midrule
                           \multicolumn{2}{r|}{\bf Total}& \bf 449  &  27 & 66     &  43   &   23  &  3   & 51  & 33 & \textbf{246} \\
                           \cmidrule{2-11}
    \end{tabular}
    \caption{Dataset of manually created MCQs. The first three columns list the names of the courses, the number of modules per course, and the number of MCQs. The remaining columns report the number of LOs from different levels of Bloom's taxonomy (RMB -- remember, UND -- understand, APP -- apply, ANL -- analyze, EVL -- evaluate, CRT -- create, N/A -- unassigned).}
    \label{tab:dataset}
\end{table*}

In contrast to the work on generating stems, current work on automated distractor generation (DG) focuses on cloze tests \cite{qiu2021automatic,jiang2017distractor,ren2021knowledge} and reading comprehension \cite{gao2019generating,zhou2020co}, as well as MCQs for domains outside of computing such as biology \cite{shin2019multiple}. These systems use the stem and/or the key to generate plausible (i.e., similar) distractors. Often, the systems generate several more distractors than necessary and select the best through a ranking system \cite{liang2018distractor}. Given that DG in technical domains poses distinct challenges,  research recognizes the potential importance of additional resources such as domain-specific ontologies \cite{kumar2023novel}. Recently, LLMs such as BERT \cite{kalpakchi2021bert} or GPT-2 \cite{offerijns2020better} have been utilized in DG. 

There is existing work on complete end-to-end MCQ generation. Rodriguez-Torrealba et al. propose an end-to-end pipeline that generates all three elements of an MCQ, with QG, QA, and DG \cite{rodriguez2022end}, using Google's T5 transformer model. They note that evaluation is difficult, only evaluating 10 questions. Leaf is another T5-powered end-to-end MCQ generation pipeline \cite{vachev2022leaf}, but has no evaluation. Unlike these systems, which focus on each task separately,  our work generates the stem, distractors, and key in a single generation step through GPT-4. Additionally, while existing systems rely on textual context to generate reading-comprehension-like questions, \cite{madri2023comprehensive,ch2018automatic} we focus on generating a question well-aligned with the provided LO, allowing for easier large-scale evaluation. Cheung et al. adopt a single-pass generation approach using ChatGPT\footnote{ChatGPT. Available at: \url{https://chat.openai.com/} [Accessed 2023-08-16]} to generate MCQs. They rely on medical textbooks to provide context \cite{cheung2023chatgpt}, noting that around 40\% of the questions are usable as-is without prompt engineering. Finally, Nasution uses the same approach to generate higher education biology MCQs without relying on a reference text, although their analysis focuses on student usage rather than instructor usability, which we focus on \cite{nasution2023using}. Our work is distinguished from \cite{nasution2023using} by rigorous prompt engineering focused on question quality and alignment, alongside a rigorous evaluation (Section \ref{sec:proposed-method}).

While the above methods focus on other areas of education, MCQs also play an essential role in computing education. Hence, there have been attempts on efficient generation of MCQs in this area as well. We could not find papers that extend Traynor and Gibson's early attempt \cite{traynor2005synthesis} to automatically generate MCQs for CS1, as it appears the focus rather shifted towards approaches based on learner-sourcing \cite{denny2008peerwise}. That is with the exception of the very recent work of Tran et al. who used GPT-3 and GPT-4 to generate isomorphic variations of pre-existing MCQs \cite{trangenerating}. Our work is most likely the first attempt to generate \emph{novel} MCQ assessments for higher education programming classes using LLMs. However, there is pre-existing work on answering MCQs with GPT-3 and GPT-4 in computing education \cite{savelka2023large,savelka2023can,savelka2023gpt}.

This has been inspired by many other instances where LLMs have already demonstrated remarkable effectiveness in generating novel language artifacts. These include course readings \cite{leiker2023prototyping}, code explanations \cite{Sarsa_2022,leinonen2023comparing,MacNeill,MacNeil2}, model solutions \cite{savelka2023thrilled,denny2022conversing,piccolo2023bioinformatics}, feedback \cite{Phung2023GeneratingHF}, or responses to help requests \cite{liffiton2023codehelp,sheese2023patterns,savelka2023efficient,kazemitabaar2023studying}. The above examples are a testament to the capabilities of the state-of-the-art LLMs in generating grammatically correct, professionally sounding language which makes the artifacts seemingly indistinguishable from the materials generated by human experts. However, such automatically generated learning resources may often lack some of the deep qualities that are necessary for them to be effective \cite{prather2023robots}. Hence, in our work we perform a rigorous evaluation of the generated MCQs.

\section{Data Set}
\label{sec:dataset}

For the experiments in this paper (Section \ref{sec:experiments}), we assembled a dataset of 246 module-level LOs coming from four Python programming\footnote{OpenEDG: Python Essentials - Part 1 (Basics). Available at: \url{https://edube.org/study/pe1} [Accessed 2023-07-30]\\ OpenEDG: Python Essentials - Part 2 (Intermediate). Available at: \url{https://edube.org/study/pe2} [Accessed 2023-07-30]\\ Coursera: Get Started with Python. Available at: \url{https://www.coursera.org/learn/get-started-with-python} [Accessed 2023-07-30]\\ Sail(): Social and Interactive Learning Platform. Available at: \url{https://sailplatform.org/courses} [Accessed 2023-07-30]} and two introductory data science\footnote{Coursera: Introduction to Data Science in Python. Available at: \url{https://www.coursera.org/learn/python-data-analysis} [Accessed 2023-07-30]\\ Coursera: Python and Pandas for Data Engineering. Available at: \url{https://www.coursera.org/learn/python-and-pandas-for-data-engineering-duke} [Accessed 2023-07-30]} courses. Three of the courses also contained MCQs (529 in total) that we collected in order to compare them to the automatically generated ones (Section \ref{sec:experiments}). We assumed that these MCQs were created manually. Two of the authors associated each MCQ with the best aligned LO from the corresponding course module. We excluded 51 of the MCQs from our experiments because we could not reasonably assign them to a single corresponding LO. An additional 29 MCQs were excluded from our study because they required more than one correct choice whereas we only focus on MCQs with single-choice keys. Hence, the resulting dataset included 449 human-crafted MCQs. Using the fine-tuned BERT classifier \cite{bertLoClassification} (Section \ref{sec:proposed-method} has details), each LO was categorized into one of the six levels of the revised Bloom's taxonomy \cite{krathwohl2002revision}. Table \ref{tab:dataset} shows the distribution of the LOs across courses and Bloom's taxonomy levels as well as the number of MCQs collected from each course. Note that some of the LOs extracted from the courses were not well-defined (e.g., no action verb) and, hence, it was not possible to categorize them in terms of Bloom's taxonomy.

\section{MCQ Generation}
\label{sec:proposed-method}
\begin{figure*}[ht]
    \centering
    \includegraphics[width=\textwidth]{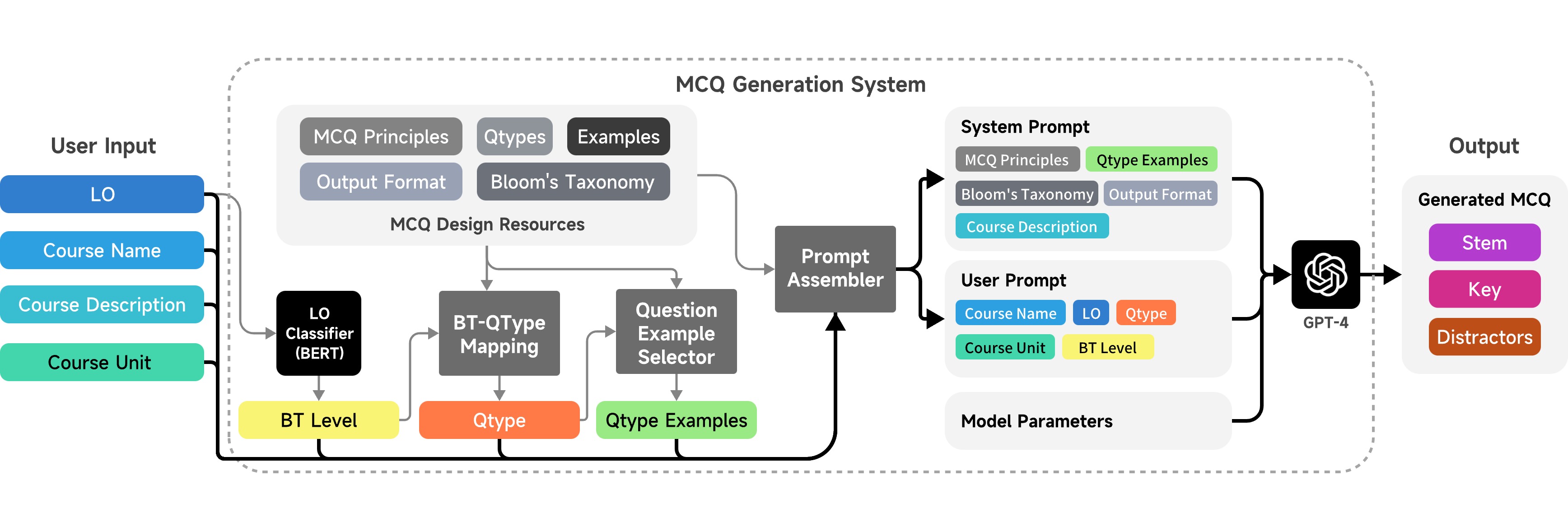}
    \vspace{-1.0cm}
    \caption{MCQ Generation Pipeline. The diagram describes the automatic MCQ generation process flowing from left to right. The input provided by the user is processed---the Bloom's taxonomy level of the provided LO is predicted which leads to determination about the appropriate type of the questions and corresponding MCQ examples. These are combined with the user input and resources from the design resources to form the system and user prompts that are submitted to the GPT-4 model outputting the generated MCQ (stem, key, distractors).}
    \label{fig:pipeline}
\end{figure*}

Figure \ref{fig:pipeline} shows the overall architecture of the MCQ generation pipeline. To generate an MCQ, we supply only high-level information about the course, the course unit (module), and the targeted LO. We then combine internal MCQ design resources with the user's input into the prompts, which are submitted to GPT-4. The below sections elaborate on each of the steps involved in the MCQ generation pipeline in greater detail.

\begin{figure*}
\centering
\begin{formal}
\scriptsize
You are a learning engineer support bot focused on creating top quality multiple-choice question assessments.\vspace{.1cm}

A multiple-choice question is a collection of three components aimed at testing a student’s understanding of a certain topic, given a particular context of what the student is expected to know. The topic, as well as the context of the topic, will be provided in order to generate effective multiple-choice questions. The three components of a multiple-choice question are as follows: a Stem, a Correct Answer, and two Distractors. There must always be only one correct answer and only two distractors.\vspace{.1cm}

The stem refers to the question the student will attempt to answer, as well as the relevant context necessary in order to answer the question. It may be in the form of a question, an incomplete statement, or a scenario. The stem should focus on assessing the specific knowledge or concept the question aims to evaluate.\vspace{.1cm}

The Correct Answer refers to the correct, undisputable answer to the question in the stem.\vspace{.1cm}

A Distractor is an incorrect answer to the question in the stem and adheres to the following properties.

\begin{enumerate}
    \item A distractor should not be obviously wrong. In other words, it must still bear relations to the stem and correct answer. 
    \item A distractor should be phrased positively and be a true statement that does not correctly answer the stem, all while giving no clues towards the correct answer.
    \item Although a distractor is incorrect, it must be plausible [...] 
    \item 4. A distractor must be incorrect. It cannot be correct, or interpreted as correct by someone who strongly grasps the topic.
\end{enumerate}\vspace{.1cm}

[...] Use ``None of the Above'' or “All of the Above” style answer choices sparingly. These answer choices have been shown to, in general, be less effective at measuring or assessing student understanding.\vspace{.1cm} 

Multiple-choice questions should be clear, concise, and grammatically correct statements. Make sure the questions are worded in a way that is easy to understand and does not introduce unnecessary complexity or ambiguity. Students should be able to understand the questions without confusion. The question should not be too long, and allow most students to finish in less than five minutes. This means adhering to the following properties.

\begin{enumerate}
    \item Avoid using overly long sentences.
    \item Avoid code that is longer than 20 lines for questions, and longer than 10 lines for the correct answer and distractors. 
    \item If you refer to the same item or activity multiple times, use the same phrase each time.
    \item Ensure that each multiple-choice question provides full context. In other words, if a phrase or action is not part of the provided topic or topic context that a student is expected to know, then be sure to explain it briefly or consider not including it.
    \item Ensure that none of the distractors overlap. In other words, attempt to make each distractor reflect a different misconception on the topic, rather than a single one, if possible.
    \item Avoid too many clues. Do not include too many clues or hints in the answer options, which may make it too obvious for students to determine the correct answer. These options should require students to use their knowledge and reasoning to make an informed choice. [...]
\end{enumerate}\vspace{.1cm}

\end{formal}

\vspace{-.45cm}

\begin{formal}[formalshade-bloom]
\scriptsize
{\bf Blooms' Taxonomy and Action Verbs:}

Multiple-choice questions must be well aligned to the learning objectives they are intended to assess students' knowledge on. This implies that they must assess skills at the right cognitive level corresponding to the Bloom’s taxonomy categorization of the learning objective. Bloom's Taxonomy offers a framework for categorizing the depth of learning, and it provides guidance on selecting appropriate action verbs when writing learning objectives. Here are the six levels of Bloom's taxonomy and their definitions:

\begin{itemize}
    \item \emph{Remember} - This level involves retrieving, recognizing, and recalling relevant knowledge from long-term memory.
    \item \emph{Understand} - At this level, learners construct meaning from oral, written, and graphic messages through interpreting, exemplifying, classifying, summarizing, inferring, comparing, and explaining.
    \item \emph{Apply} - This level requires learners to carry out or use a procedure through executing or implementing it.
    \item \emph{Analyze} - At this level, learners break material into constituent parts, determine how the parts relate to one another and to an overall structure or purpose through differentiating, organizing,[...]
    \item \emph{Evaluate} - This level involves making judgments based on criteria and standards through checking and critiquing.
    \item \emph{Create} - At this level, learners put elements together to form a coherent or functional whole, or they reorganize elements into a new pattern or structure through generating, [...]
\end{itemize}
\end{formal}

\vspace{-.45cm}

\begin{formal}[formalshade-course]
\scriptsize
{\bf Course Context}

Below is a brief description of the \textcolor{data}{Practical Programming with Python} course.\vspace{.1cm}

Description: \textcolor{data}{Students learn the concepts, techniques, skills, and tools needed for developing programs in Python. Core topics include types, variables, functions, iteration, conditionals, data structures, classes, objects, modules, and I/O operations. Students get an introductory experience with several development environments, including Jupyter Notebook, as well as selected software development practices, such as test-driven development, debugging, and style. Course projects include real-life applications on enterprise data and document manipulation, web scraping, and data analysis.}
\end{formal}

\vspace{-.45cm}

\begin{formal}[formalshade-qtype]
\scriptsize
{\bf Question type: \textcolor{data}{Recall}}

\color{data}

A recall multiple-choice question often contains minimal code, if at all, in its stem. They may assess a student's understanding of basic programming concepts or include some technical details. It should be conceptual while containing specific knowledge of the course content and learning objectives.\vspace{.1cm}

In the context of an Introductory Programming with Python course, these questions typically ask about Python syntax and principles, built-in functions, or standard libraries. They may also evaluate students’ understanding of fundamental programming concepts such as coding conventions and object-oriented programming (OOP) principles.\vspace{.1cm}

Below are some examples:

Example 1:
\begin{verbatim}
{
    "question": "Which of the following methods can be used to remove a single element from a list in Python?",
    "choices": [{"option": "A", "text": "pop()"}, {"option": "B", "text": "delete()"}, {"option": "C", "text": "clear()"}],
    "correctAnswer": "A",
    "explanation": "clear() will remove all elements, you can use del but not delate() to remove element."
} [...]
\end{verbatim}
\end{formal}

\vspace{-.45cm}

\begin{formal}[formalshade-output]
\scriptsize
{\bf Output Format}

Output your multiple-choice question in an easy-to-parse json dictionary format, where the stem is the key, and the correct answer and distractor choices are values. Be sure to clearly distinguish which choice is the correct answer and which are distractors. The question generated should have exactly 2 distractors and 1 correct answer (3 choices in total). If there is code in the stem, please set ``code\_in\_stem'' to True. If there is no code in the stem, set ``code\_in\_stem'' to False.\vspace{.1cm}

Your return should be the exact json structure of the following example: 

\begin{verbatim}
{
    "question": "The stem of the question", 
    "choices": [{"option": "A", "text": "Answer Choice A in string type"}, {"option": "B", "text": "Answer Choice B in string type"}, 
                {"option": "C", "text": "Answer Choice C in string type"}],
    "correctAnswer": "A",
    "code_in_stem": "True or False",
    "explanation": "The explanation of the choices"
}  
\end{verbatim}

[...] If any of the multiple-choice items contain code, please format the code snippet as shown below:

\begin{verbatim}
```python
def test():
    return "Correct Format"
```
\end{verbatim}
\end{formal}

\begin{textblock*}{6.2cm}(13.75cm,-8.93cm)
\includegraphics[width=.28\textwidth]{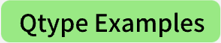}
\end{textblock*}

\begin{textblock*}{6.2cm}(13.8cm,-20.0cm)
\includegraphics[width=.265\textwidth]{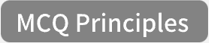}
\end{textblock*}

\begin{textblock*}{6.2cm}(13.65cm,-13.2cm)
\includegraphics[width=.31\textwidth]{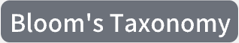}
\end{textblock*}

\begin{textblock*}{6.2cm}(13.82cm,-5.15cm)
\includegraphics[width=.255\textwidth]{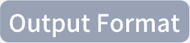}
\end{textblock*}

\begin{textblock*}{6.2cm}(13.6cm,-10.5cm)
\includegraphics[width=.33\textwidth]{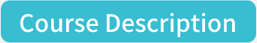}
\end{textblock*}

\caption{The System Part of the Prompt. The figure shows extensive excerpts from the system part of the prompt showing the main constituents: MCQ Principles, Bloom's Taxonomy, Course Description, Question Type Examples, and Output Format. The colored stripes on the left and the colored badges match the colors of the pipeline constituents from Figure \ref{fig:pipeline}. The \texttt{[...]} tokens mark places where the text has been abridget to fit on the page. The purple text is dynamic (data dependent).}
\label{fig:system_prompt}
\end{figure*}

\begin{figure}
\centering
\scriptsize
\begin{formal}[formalshade-qtype2]
Generate a top quality \textcolor{data}{Recall} multiple-choice question
\end{formal}

\vspace{-0.25cm}

\begin{formal}[formalshade-cname]
for the course \textcolor{data}{Practical Programming in Python}
\end{formal}

\vspace{-0.25cm}

\begin{formal}[formalshade-unit]
on the unit: \textcolor{data}{Python Basics and Introduction to Functions}.
\end{formal}

\vspace{-0.25cm}

\begin{formal}[formalshade-lo]
Your generated question should target the following learning objective: 

\textcolor{data}{Explain what Python is and how to use it to run single-line expressions as well as\\ small multi-line programs.}
\end{formal}

\vspace{-0.25cm}

\begin{formal}[formalshade-btlevel]
Your generated question should also be at the \textcolor{data}{understand} level in Bloom's\\ taxonomy.
\end{formal}
\begin{textblock*}{6.2cm}(4.90cm,-3.55cm)
\includegraphics[width=.14\textwidth]{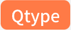}
\end{textblock*}
\begin{textblock*}{6.2cm}(4.55cm,-3.0cm)
\includegraphics[width=.25\textwidth]{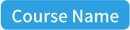}
\end{textblock*}
\begin{textblock*}{6.2cm}(4.65cm,-2.5cm)
\includegraphics[width=.22\textwidth]{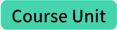}
\end{textblock*}
\begin{textblock*}{6.2cm}(5.08cm,-2.0cm)
\includegraphics[width=.08\textwidth]{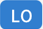}
\end{textblock*}
\begin{textblock*}{6.2cm}(4.7cm,-1.1cm)
\includegraphics[width=.2\textwidth]{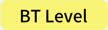}
\end{textblock*}
\caption{The User Message Part of the Prompt. The figure shows an example user message part of the prompt showing the main constituents: Question Type, Course Name, Module (unit) Name, Learning Objective, and Bloom's Taxonomy Level. The colored stripes on the left and the colored badges match the colors of the pipeline constituents from Figure \ref{fig:pipeline}. The purple text is dynamic (data dependent).}
\label{fig:user_prompt}
\end{figure}

\subsection{User Input}
The notable feature of the proposed MCQ generation pipeline is that the user is expected to provide only high-level information about the course and the module. This is in stark contrast to the large majority of other MCQ generation systems described in Section \ref{sec:related_work}. While those systems use a piece of text (e.g., a paragraph from a textbook) to generate an MCQ, our system utilizes a specific module-level LO to generate an MCQ that is well-aligned with that LO. This allows us to more carefully align MCQ generation with assessment of student achievement of the intended LO. Besides the specific module-level LO and the module it comes from, the user is expected to provide a course title (see Table \ref{tab:dataset} for titles of the courses used in this study), a short course description with the list of course-level LOs, and the list of course modules. Using this particular context as well as following the best practices for prompting an LLM \cite{nasution2023using,cheung2023chatgpt} to generate an MCQ are the prominent features of the presented pipeline. 

\subsection{Design Resources}
We curated a set of static resources to support various stages of effective MCQ generation. We focused on the following elements:

\begin{itemize}[leftmargin=*]
    \item \emph{MCQ principles} -- A set of research-validated and generally accepted principles that guide authoring of high-quality MCQs. For example, this includes that distractors should be plausible and limited in number (often just two). We only provided a concise description here but the \emph{MCQ Principles} section of Figure \ref{fig:system_prompt} shows an extensive excerpt.
    \item \emph{Bloom's taxonomy} -- Definitions of the six levels of the revised Bloom's taxonomy \cite{krathwohl2002revision}, i.e., remember, understand, apply, analyze, evaluate, and create. This taxonomy helps educators articulate LOs that focus on concrete actions and behaviors, and target distinct levels of cognitive processes. For LOs to guide the selection of assessments, they must be measurable, i.e., it should be possible to evaluate whether learners attained the intended LO. We also include information on the aims and uses of Bloom's taxonomy. The \emph{Bloom's taxonomy} section of Figure \ref{fig:system_prompt} shows an extensive excerpt.
    \item \emph{Question type system} -- From an informal analysis of the collected dataset of human-crafted MCQs (Section \ref{sec:dataset}), we define five types of programming MCQs: recall, fill-in the blank, identify correct output, analyze (trace) code, and scenario (see the \emph{LO Mapping to Question Types} step of the pipeline below). We include the definitions of these types into the resources. The \emph{QType Examples} section of Figure \ref{fig:system_prompt} provides an example of the recall type.
    \item \emph{Question type examples} -- A small set of high-quality MCQ examples for each of the defined types (example in the \emph{QType Examples} section of Figure \ref{fig:system_prompt}). 
    \item \emph{Output format} -- Specification of the output format (JSON). Figure \ref{fig:system_prompt} shows additional details (\emph{Output Format} section).
\end{itemize}

\noindent Providing an LLM with this type of information to create effective MCQs is one of the core contributions of this work. Unlike Leaf \cite{vachev2022leaf}, our methodology allows for training-free generation of questions and distractors, ensuring effective generation without necessarily worrying about out-of-domain generalization. It is reasonable to assume that this holds for mainstream domains, such as introduction to programming, where plenty of related materials were included in the dataset used for pre-training of the LLM. However, this approach may not be directly applicable in highly specialized domains with less abundant content.


\subsection{LO Bloom's Taxonomy Level Classifier}
We fine-tuned a binary BERT classifier~\cite{bertLoClassification} for each Bloom’s Taxonomy
category on 21,380 LOs from 5,558 university courses. BERT (bidirectional encoder representation from transformers) \citep{devlin2018bert,vaswani2017attention} is a popular LLM notable for its fine-tuning capabilities on a down-stream task. We used the models to predict the Bloom’s Taxonomy level (i.e., remember, understand, apply, analyze, evaluate, or create) of the generated LOs. The prediction is then utilized in the \emph{LO Mapping to Question Types} step of the pipeline (see below), and it is also embedded in the user message part of the prompt as shown in Figure \ref{fig:user_prompt}.

\subsection{LO Mapping to Question Types}
We use the automatically predicted Bloom's taxonomy level of the provided LO (see above) to determine the appropriate MCQ types for the LO (see Table \ref{tab:BT_mapping}). We map each taxonomy level to one or more question types and a question is generated for each type:

\begin{itemize}
    \item \emph{Recall} -- Given minimal to no code, students are asked about basic programming concepts or technical details.
    \item \emph{Fill in the Blank} -- Given a code snippet with some sections removed, students are asked to select an option that successfully replaces the blank to create syntactically and semantically correct code.
    \item \emph{Scenario Based} -- Given a scenario or situation, students are asked to identify appropriate tools, methods, or packages best suited to accomplish the task prescribed.
    \item \emph{Correct Output} -- Given a code snippet, students are asked to trace program execution to determine either intermediate or final outputs.
    \item \emph{Code Analysis} -- Given a code snippet, students are asked to spot errors or to build on or use the code in a new manner.
\end{itemize}

\noindent These question types allow us to target specific levels of cognitive processes, e.g., preventing the generation of MCQs involving complex code tracing for LOs asking students to remember a concept.


Finally, high-quality MCQ examples are selected based on the question type in order to ensure that the proper cognitive processes are targeted in accordance with the provided LO. The corresponding examples of the high-quality MCQs of that type are retrieved from the design resources.

\subsection{Prompt}
GPT-4 utilizes the so-called system part of the prompt to provide the overall context of the conversation. From the design resources, we retrieve a set of MCQ principles and best practices. We provide the definition of the predicted Bloom's taxonomy level, examples of the selected question type, the course description, and the expected output format of the generated MCQ as well. Figure \ref{fig:system_prompt} shows an extensive example of the system part of the prompt.

The user message part of the prompt is the piece of text to which the model is expected to respond (within the context established by the system part of the prompt). We include the user-provided course name and the course module, the specific module-level LO, the predicted Bloom's taxonomy level, and subsequently mapped question type into the user prompt. Figure \ref{fig:user_prompt} shows an extensive example of the user message part of the prompt.

\subsection{MCQ Generation Step}
We use GPT-4 (\verb|gpt-4-0613|), which is one of the most advanced LLMs released by OpenAI as of the writing of this paper \cite{openai2023gpt4}.  As for the model parameters, we set \texttt{temperature} to its default value of 1.0. We keep \texttt{top\_p}, \texttt{frequency\_penalty}, and  \texttt{presence\_penalty} at their default values of 1.0, 0.0, and 0.0 respectively. We impose a \texttt{max\_token} limit of 2,000 tokens (i.e., the maximum length of a generated MCQ). 

As the response to the prompt, consisting of the system part and the user message part, the GPT-4 model generates an output in the expected JSON format. The response contains the stem, the key, and the distractors, i.e., all the needed constituents of an MCQ. One of the choices is marked as correct.

\begin{figure*}[t]
    \centering
    \includegraphics[width=\textwidth]{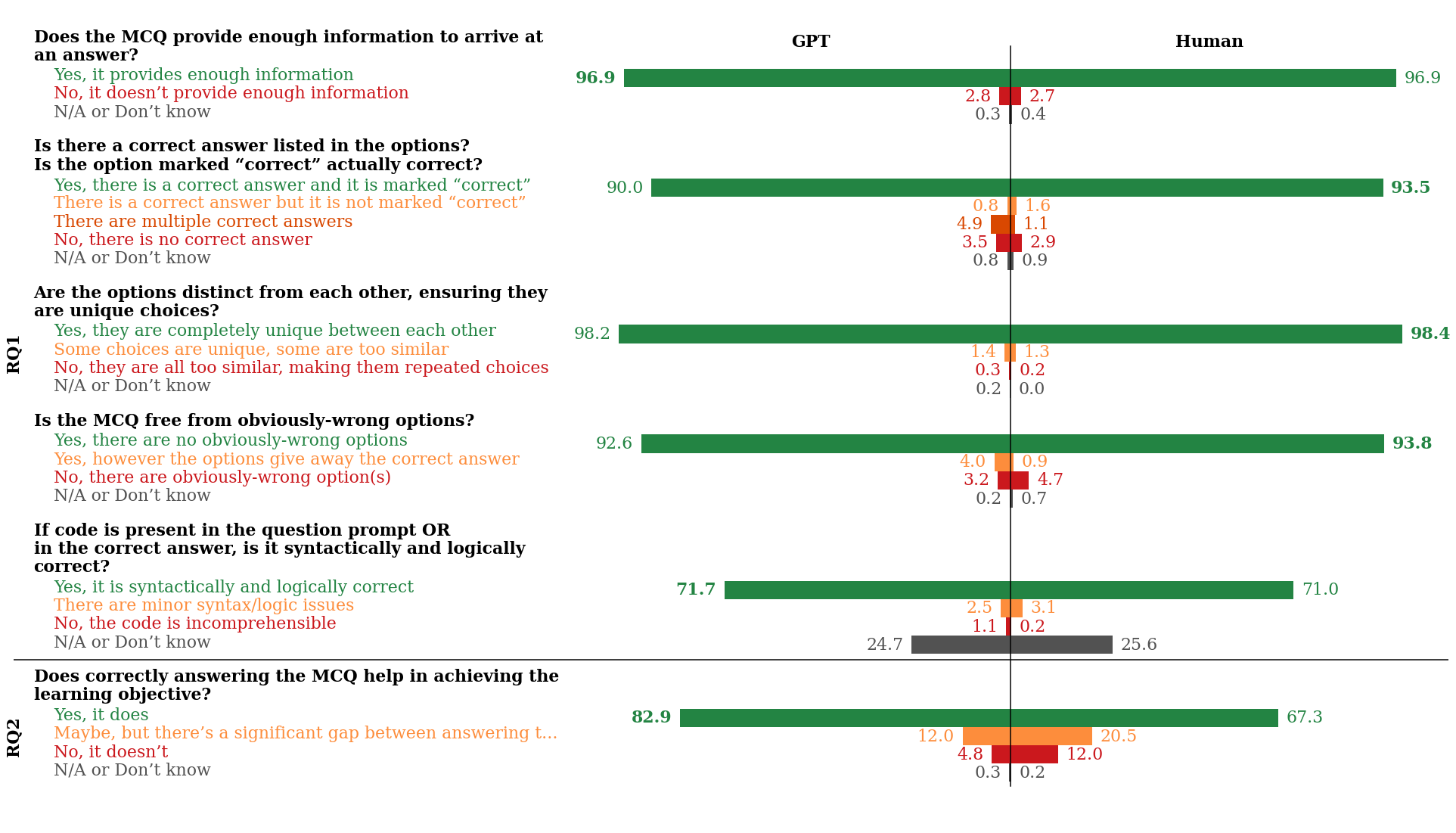}
    \vspace{-1.2cm}
    \caption{MCQ Evaluation Results. The MCQ evaluation rubric is shown on the left. On the right, the automatically generated MCQs (GPT) are compared with the human-crafted ones. The top part of the figure shows the results for the five rubric items focused on the quality of the MCQs (RQ1). The bottom part shows the rubric item addressing the LO-MCQ alignment (RQ2). The reported ratios are based on the counts of the labels after the disagreement resolution described in Section \ref{sec:experiments}.}
    \label{fig:results}
\end{figure*}

\section{Experimental Design}
\label{sec:experiments}
To evaluate the quality of the automatically generated MCQs, we generate several MCQs for each of the 246 LOs in our dataset. We generate a single question per question type mapped to the LO's Bloom's taxonomy level as described in Table \ref{tab:BT_mapping}. When we could not assign a Bloom's taxonomy level we simply generate MCQs of all types. Following this process we obtain a sizeable dataset of 651 automatically generated MCQs. Table \ref{tab:BT_mapping} shows the distribution of the MCQs per both their question type and Bloom's taxonomy level.

\begin{table}
    \centering
    \begin{tabular}{c@{\hspace{4pt}}c@{\hspace{4pt}}c@{\hspace{4pt}}c@{\hspace{4pt}}c@{\hspace{4pt}}c@{\hspace{4pt}}c@{\hspace{4pt}}c@{\hspace{4pt}}c@{\hspace{4pt}}}
        MCQ-Type & RMB & UND & APP & ANL & EVL & CRT & N/A & Total\\
        \hline
        Recall & 27 & 66 & -  & -  & - &  - &32 & 125\\
        Fill in the Blank & 27 & 63 & 41 & 23 & - & - &32 &186 \\
        Scenario Based & - & -& 41 & 21 & - & - &31 & 93\\
        Correct Output & - & - & 41 & 23 & 3 & - &33 & 100\\
        Code Analysis & - & - & 41 & 21 & 3 & 49 &33 &147\\
        \hline 
        Total & 54 & 129 & 164 & 88 & 6 & 49 & 161 & \textbf{651}\\
        \hline 
    \end{tabular}
    \caption{Automatically Generated MCQs. The table shows the distribution of the automatically generated MCQs per question types (rows) and the levels of Bloom's taxonomy (columns).}
    \label{tab:BT_mapping}
\end{table}

We developed a rubric consisting of six criteria shown in Figure \ref{fig:results}. The first five criteria of the rubric target RQ1, while the last criterion targets RQ2. Seven students and six CS instructor annotators (all authors of this paper) applied the rubric to the generated (651) as well as human-crafted (449) MCQs, i.e., 1,100 MCQs in total.


\begin{table}[t]
    \centering
    \begin{tabular}{lcc}
    \toprule
    \bf Rubric item            &\bf Gwet's AC1 &\bf Fleiss $ \kappa $ \\
    \midrule
    Sufficient information     & 0.96           & 0.35 \\
    Correct answer             & 0.88           & 0.31 \\
    Unique choices             & 0.62           & 0.16 \\
    No obviously wrong choice  & 0.92           & 0.07 \\
    Correct code               & 0.88           & 0.07 \\
    LO alignment               & 0.93           & 0.23 \\
    \bottomrule
    \end{tabular}
    \caption{Inter-rater Agreement. The table shows Gwet's AC1 and Fleiss $\kappa$ scores for each of the six rubric items (Figure \ref{fig:results}).}
    \label{tab:irr}
\end{table}

Each annotator was asked to complete 250 annotations\footnote{Due to time, most annotators completed over 240 annotations, with one only completing 93}. We required annotators to attempt answering the question before applying the rubric items. The overall inter-rater agreement in terms of Fleiss $\kappa$ was 0.22 which corresponds to a fair agreement. The $\kappa$ statistic is known to severely underestimate the agreement in situations when one of the labels is clearly dominant \cite{zec2017high}. This is the case of our data where for each rubric item there is always a clearly dominant answer. To address this issue we also compute Gwet's AC1 which is robust in such situations \cite{wongpakaran2013comparison}. Table \ref{tab:irr} reports the AC1 and $\kappa$ scores for each of the six rubric items. The measured Gwet's AC1 range from 0.62 to 0.96. Each MCQ was annotated by at least one student and one instructor. Overall, 3,076 annotations for the 1,100 MCQs were produced, i.e., a little less than 3 annotations per MCQ on average. To resolve disagreements we used the following rules in the presented order: 

\begin{enumerate}
    \item Majority Vote; 
    \item Instructors' annotations had precedence over students';
    \item An annotator who correctly answered the MCQ had precedence over the one who answered incorrectly;
    \item The least favorable evaluation. 
\end{enumerate}

\noindent Then, the automatically generated MCQs were compared to the human-crafted ones in terms of the six categories from the developed rubric. To test for statistically significant differences we employ Fisher's exact test (extended to multiple categories through approximation).

\section{Results}

\subsection{MCQ Quality (RQ1}
Figure~\ref{fig:results} shows the results of the experiments described in Section \ref{sec:experiments}. The first five criteria relate to RQ1. From those, we observe that the generated MCQs appear to be of comparable quality to the human-crafted ones. A notable issue is that 4.9\% of the automatically generated MCQs had multiple correct answer choices (compared to only 1.1\% of the human-crafted MCQs). The quality of the distractors appears to be another rubric item where the MCQ generation pipeline struggles. GPT MCQs were more likely to be annotated as having distractors which gave away the correct answer (4.0\% vs 0.9\% human). We performed the Fisher exact test to test for statistical significance. The presence of the correct answer ($p=0.002$) and the presence of obviously wrong options ($p=0.002$) criteria had a statistically significant difference between automatic and human generated questions. This corresponds to the above described differences related to the presence of multiple correct choices in the automatically generated MCQs and distractors which gave away the correct answer. Hence, we conclude that the generated MCQs provide sufficient information in clear language (RQ1.i) and contain syntactically and logically correct code (RQ1.iv). That is they do not appear to differ from the human-crafted MCQs in terms of the mentioned quality criteria. On the other hand, the generated MCQs are somewhat lacking when compared to the human-crafted ones when it comes to having a single correct answer (RQ1.ii) and high-quality distractors (RQ1.iii).

\subsection{MCQ-LO Alignment RQ2}
The results of evaluating RQ2 are also shown in Figure \ref{fig:results}---the last rubric item. We observed that the automatically generated MCQs appear to be noticeably better aligned with the LOs as compared to the human-crafted ones. This is confirmed by the Fisher exact test as well ($p<10^{-9}$). It appears that human generated MCQs were quite often related to the LOs but did not target the appropriate cognitive level (Bloom's taxonomy) or simply were too different to be considered a viable assessment for the LO (20.5\% vs 12.0\% auto-generated). Often, the human-crafted MCQs did not relate to the LOs at all (12.0\% vs 4.8\% auto-generated). This finding is discussed in great detail below.

\section{Discussion}


Generated MCQs were more likely to have multiple correct answers. This was mostly observed in MCQs generated for the LOs at the \emph{Apply} and \emph{Create} cognitive levels of Bloom's Taxonomy. Consider the example shown in Figure \ref{fig:correct_answer_example}. While the answer choices comprise three different methods of converting a string to a list and reversing it, all three of them accomplish the same task with no side-effects. Hence, all of the options are correct. Note that only the first one was marked as correct by GPT-4. While not too prevalent (4.9\% of generated MCQs), this issue is serious and requires human intervention to be fixed. Future work should focus on mitigating it.

\begin{figure}
\centering
\begin{formal}[formalshade-unit]
Given the string \texttt{s = "Hello, World!"} you need to create a piece of code that will return a list of the words in \texttt{s} but in reversed order while preserving the original string. Which one of the following code snippets achieves this goal? \\

A. reversed\_s = s.split(' ')\\  \-\hspace{.3cm} reversed\_s.reverse()\\
B. reversed\_s = ''.join(reversed(s)) \\ \-\hspace{.3cm} reversed\_s = reversed\_s.split(' ')\\
C. reversed\_s = s.split(' ')[::-1]
\end{formal}
\caption{An example generated MCQ where all the choices are correct answers. Only the first choice (A.) was marked as correct by GPT-4.}
\label{fig:correct_answer_example}
\end{figure}

Compared to the human-crafted ones, the generated MCQs were more likely to have obviously-wrong choices. This was the most pronounced in the \emph{Fill-in-the-Blank} and \emph{Scenario Based} MCQ type. Of the MCQs that were annotated as having options that ``give away the correct answer'', the majority gave the answer away in the question stem. This issue is somewhat less pressing than the one described earlier as it does not render the MCQ completely invalid. While ideally an instructor would recognize the issue and edit the MCQ this cannot be relied on. Figure \ref{fig:give_away_example} shows an example MCQ where the code snippet included in the stem gives away the correct answer.

\begin{figure}
\centering
\begin{formal}[formalshade-lo]
In Python, the \_\_\_\_\_ loop is used when we want to iterate over a sequence (like a list, tuple, set, or string) or other iterable objects. Iterating over a sequence is called traversal.
\begin{verbatim}
for item in iterable:
    # execute some statements
\end{verbatim}
A. for\\
B. while\\
C. do\\
\end{formal}
\caption{An example generated MCQ where the code snippet included in the stem gives away the correct answer (A).}
\label{fig:give_away_example}
\end{figure}



GPT-4's ability to produce effective MCQs was close to human performance. In MCQs where GPT failed to meet a quality requirement this was usually the only issue with the question. Based on our analysis, the most serious issue with the generated MCQs appear to be implausible or plainly incorrect distractors along with revealing the correct answer in the question stem. Future work should focus on prompt engineering techniques to mitigate these issues. Importantly, 81.7\% of all the generated MCQs passed all of the evaluation criteria. This suggests that less than 1 in 5 generated questions would require instructors' edits.

The proposed pipeline is designed to generate a single MCQ for a single LO, and in the vast majority of cases, the MCQ was well-aligned with the LO. We observed that the alignment of the generated MCQs was vastly superior compared to the alignment of the human-crafted ones (e.g., Figure \ref{fig:lo_align_example}). This is likely due to the fact that educators may often focus more on the alignment of an MCQ with a module's topic and less on the alignment with an LO. Achieving the alignment between LOs and assessments is often challenging for educators. This provides an excellent motivation for a support system such as the one presented in this paper.

\begin{figure}
\centering
\begin{formal}[formalshade-btlevel]
{\bf Learning Objective}\\
Discuss the importance of writing comments and how to write correct comments.\\

\ 

{\bf Generated MCQ}\\
In Python programming, why is it important to write comments in your code?\\
A. Comments are functional elements of the code that can affect program execution.\\
B. Comments help to improve the readability and maintainability of the code by explaining the function and intention of parts of the code.\\
C. Comments are used to debug if there is an error in the program execution.\\

\ 

{\bf Human-crafted MCQ}\\
Clean code is easy to maintain.\\
A. True\\
B. False
\end{formal}
\caption{An example where the generated MCQ is better aligned with the LO than the human-crafted one which only appears to target the module's topic and not the LO itself.}
\label{fig:lo_align_example}
\end{figure}

During our analysis, we considered Bloom's Taxonomy, MCQ-type, LO, and course title as potential factors for MCQ-LO alignment. We observed that relative to the other courses, MCQs generated from the LOs in the \textit{Python Essentials 1} course scored slightly worse on alignment. We found that the failure cases for GPT MCQ alignment occurred independently of these variables. While our results are promising, there are some caveats. As we scraped MCQs from the internet, we needed to manually associate MCQs with LOs. This likely explains some part of the observed difference in MCQ-LO alignment between the generated and human-crafted MCQs. 

When evaluating the MCQs the human raters were asked to answer the questions. While such a setup does not replace the evaluation of the questions in real classroom settings it may provide tentative insights into the difficulty of the generated MCQs compared to the human-crafted ones. Overall, the human-crafted MCQs appear to be more challenging than the generated ones. The student raters answered 71.5\% of the generated MCQs correctly (62.6\% human-crafted) while the instructor raters handled 80.1\% of the generated questions successfully (76.3\% human-crafted). It is important to emphasize that the raters were not instructed to put effort into answering the questions correctly. Hence, the reported rates need to be treated with caution.

\section{Implications for Teaching Practice}
Our results suggest that LLM-powered tools can generate MCQs that have comparable quality to MCQs generated by humans while achieving better alignment with LOs. Programming instructors can use LLM-powered tools to reduce their workload, enabling them to focus more on student engagement and curriculum enhancement. Deploying such tools could make updating and revising assessments more efficient. 


Of particular note is the finding that automatically generated MCQs seem to be better aligned with the LOs than those generated by humans. This could mean that use of automatically generated MCQs might provide more accurate representations of students' mastering of the intended LOs, potentially leading to more accurate evaluations. Moreover, it indicates that LLM-powered tools could be used not only for assessment generation but also as an invaluable tool for curriculum design and planning \cite{sridhar2023harnessing}. 


Given the potential of automated MCQ generation, teaching practice could benefit enormously from incorporating the LLM-based tools into the assessment design process. As with the introduction of any novel tool, care should be taken to correctly deploy and use such resource, and to handle its output responsibly (e.g., editing generated MCQs that have issues).

\section{Limitations and Threats to Validity}

While we attempted to make the assessment of the quality of automatically generated MCQs as objective as possible via a well-defined rubric, we acknowledge potential bias from the limited pool of human raters. Additionally, the human generated questions were created before the study and, hence, the instructors authoring the MCQs were not aware of the rubric that we used to evaluate the questions in this study. It is plausible that if the rubric was available to them they could have authored the MCQs to better satisfy the requirements. Finally, as we manually paired human-crafted MCQs to LOs, it is possible we ourselves caused part of the observed MCQ-LO misalignment.

Our study has not evaluated the utility of auto-generated MCQs in live classrooms, and, hence, their pedagogical impact is not known. Although, the LLM-powered system was able to generate questions that seem comparable to those written by humans, it remains to be seen how effective these auto-generated MCQs are at assessing student learning. Additionally, we did not compare the difficulty of the generated MCQs to that of the human-crafted ones. We focused on the alignment between an MCQ and a corresponding LO. However, we did not consider the alignment between an MCQ and the learning content. These important investigations are left for future work.

Our research focused only on Python programming courses at the higher education level, which may limit the generalizability of the results to other programming languages or education levels. While Python is widely used in introductory programming courses, it is possible that LLMs may perform differently when generating MCQs for other languages or more specialized domains. Since the evaluated approach relies on the LLM's ``knowledge'' of the domain it is quite likely that it would not generalize well to highly specialized domains.


\section{Conclusions and Future Work}
This paper provides promising evidence suggesting that systems powered by LLMs can automatically generate high-quality MCQs for Python programming courses. The generated MCQs are comparable in quality to those created by human educators (RQ1) and exhibit strong alignment with LOs (RQ2). Overall, this study demonstrates the feasibility of high-quality automated MCQ generation that has the potential to significantly reduce the time and effort educators currently spend on developing assessments


Future work needs to explore the effectiveness of LLM-generated MCQs within real-world classroom contexts. This will allow a better understanding of the uptake, utility and discriminative power of generated MCQs. Additionally, future work needs to be done to better understand the novelty and diversity of  MCQs generated with LLMs. Currently, it is not easy to understand how capable this system is at generating entire quizzes and entire quiz pools for particular LOs. Furthermore, we do not assess the refinement capabilities of GPT-4 for this task. Given that GPT-4 is a chat completion model, providing further feedback should lead to enhanced generation, and we do not study these capabilities. There is also the need to understand the effects of different parameter settings, such as temperature \cite{agarwal2023understanding}. Lastly, extending this work to other programming languages and determining the effects on MCQ quality and alignment with LOs would be an important avenue to explore.


\bibliographystyle{ACM-Reference-Format}
\bibliography{main}

\end{document}